# WEB OBJECT SIZE SATISFYING MEAN WAITING TIME IN MULTIPLE ACCESS ENVIRONMENT

Y. –J. Lee

Department of Technology Education, Korea National University of Education, Cheongju, South Korea

*ABSTRACT*

*This paper addresses web object size which is one of important performance measures and affects to service time in multiple access environment. Since packets arrive according to Poission distribution and web service time has arbitrary distribution, M/G/1 model can be used to describe the behavior of the web server system. In the time division multiplexing (TDM), we can use M/D/1 with vacations model, because service time is constant and server may have a vacation. We derive the mean web object size satisfying the constraint such that mean waiting time by round-robin scheduling in multiple access environment is equal to the mean queueing delay of M/D/1 with vacations model in TDM and M/H$_2$/1 model, respectively. Performance evaluation shows that the mean web object size increases as the link utilization increases at the given maximum segment size (MSS), but converges on the lower bound when the number of embedded objects included in a web page is beyond the threshold. Our results can be applied to the economic design and maintenance of web service.*

## 1. INTRODUCTION

Simultaneous access of multiple users to a server in the web environment increases the mean waiting delay of an end-user. Therefore, the quality of service (QoS) degradation problem of the end-user arises. In order to develop a technology for solving this problem, we first should find the mean waiting delay of the end-user accurately.

Generally, the user's request to the web server per unit time follows the Poisson distribution and the web service time follows the general distribution instead of the Exponential distribution. M/G/1 model [1] is known to be suitable to describe a web service. In particular, because web services are influenced by the size of the web objects, Shi et al [2] presented the result that as a statistical distribution to describe the web service, Weibull distribution and Exponential distribution are suitable. Meanwhile, Khayari et. al [3] and Riska et. al [4] have presented an algorithm to fit the empirical distribution to the Hyper-exponential distribution. When the web service is given by the Hyper-exponential distribution in the steady state, the research [5] to obtain the number of concurrent users satisfying the average queueing delay was conducted. However, more empirical researches for web services distribution related to the Internet are still needed. Additionally, M/G/1 with vacations model [6, 7, 8, 9] as a modification of M/G/1 model has proposed.

In the time division multiplexing (TDM), time quantum assigned to each user is slotted so that the data transmission takes place just only at the starting point of the slot. Therefore, if the system is empty at the beginning of the slot, the server goes to the vacation state during that time slot. To apply the TDM scheme to the queueing system can be described as an M/D/1 with vacations model in which the service distribution is given by the constant.

On the other hand, when several users simultaneously request a web object of the packets in the web server, and the round-robin scheduling for the web service is used, we can determine the mean waiting time. When the system is in the steady state, we can infer that the mean waiting time is equal to the mean queueing delay by the M/D/1 with vacations model in the TDM approximately.

The objective of this study is to find the web object size satisfying that the mean waiting time for multiple web access environments is equal to the mean queueing delay for M/D/1 with vacations model in TDM and the mean queueing delay for M/H$_2$/1 model, respectively. We first find the number of simultaneous users satisfying that M/H$_2$/1 queueing delay is equal to the queueing delay in TDM. And then we obtain the web object size. The reason to obtain the web object size satisfying delay constraint of end-user is why the controlling of that is the most economic way in the design and operation of the web service.

The rest of this paper is structured as follows. In the next section, we discuss the M/D/1 with vacations model based on Modiano [8] and Bose [6] in TDM and the M/H$_2$/1 queueing model. In section 3, we first describe the model to find the mean waiting time by round-robin scheduling in the multiple web access environment. We then determine the web object size satisfying the constraint that the mean waiting time equal to the mean queueing delay for M/D/1 with vacations model in TDM and M/H$_2$/1 model, respectively. In section 4, we present and analyze the performance evaluation results. Finally, in section 5, we discuss the conclusions and future research.

## 2. QUEUEING DELAY FOR M/D/1 WITH VACATIONS MODEL AND M/H$_2$/1 MODEL

### 2.1. Queueing Delay for M/D/1 with vacation Model

We consider a single-server queueing system where object requests arrive according to a Poission process with rate λ, but service times have a general distribution (M/G/1). By Pollaczek-Khinchin formula, the expected mean queueing delay is given by (1) [1, 10].

$$W = \frac{\lambda E(S^2)}{2(1-\rho)} \quad (1)$$

where $\rho=\lambda/\mu= \lambda E(S)$. *S* is the random variable representing the service time and identically distributed, mutually independent, and independent of the inter-arrival times.

If the service times are identical for all requests (M/D/1), that is $E(S^2) = 1/\mu^2$, equation (1) becomes (2).

$$W = \frac{\rho}{2\mu(1-\rho)} \quad (2)$$

Now, we consider M/G/1 queueing model with vacations [9]. In this model, when the queue is empty, the server's vacation time is represented as IID random variable, and independent with service times and arrival times. When the system is empty after the vacation, the server can take another vacation. In this model, the arrival distribution at any time t, should be the same as the length of the queue [7], *R* is given by (3)

$$R = \frac{\lambda E(S^2)}{2} + \frac{(1-\rho)E(V^2)}{2E(V)} \tag{3}$$

Where, $E(V)$ and $E(V^2)$ are the primary and secondary moment of vacation time, respectively. From (1), the mean queuing delay of M/G/1 with vacations model is derived as (4).

$$W = \frac{R}{1-\rho} = \frac{\lambda E(S^2)}{2(1-\rho)} + \frac{E(V^2)}{2E(V)} \tag{4}$$

Now, consider the time division multiplexing (TDM) system such as Figure 1.

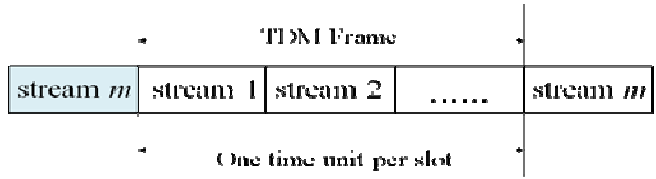

Figure 1. TDM system

Figure 1 shows that the *m* fixed-length packets with each $\lambda/m$ arrival rate are multiplexed and arrive into the system according to the Poisson distribution. Total traffic is $\lambda$, the service rate ($\mu$) is equal to $1/m$, and the load on the entire system is $\rho = \lambda$. Thus, equation (2) with $\mu=1/m$ and $\rho = \lambda$ gives the mean queuing delay per packet as (5). This delay represents time in the frequency division multiplexing (FDM) [10]. Equation (5) can be also obtained by setting $E(S) = E(S^2) = 1/\mu = m$, and $E(V) = m$, $E(V^2) = 1/\mu^2 = m^2$ in (4).

$$W_{FDM} = \frac{\rho m}{2(1-\rho)} \tag{5}$$

In the TDM, where *m* traffic streams are time division multiplexed in a scheme, whereby the time axis is divided in *m*-slot frames with one slot dedicated to each traffic stream (Figure 1). Thus the mean queueing delay in TDM is given by (6) [10].

$$W_{TDM} = W_{FDM} + \frac{m}{2} = \frac{m}{2(1-\lambda)} \tag{6}$$

## 2.2. Queueing Delay for M/H$_2$/1 Model

Generally, web objects are composed of two types: static and dynamic. A static object is one home page first requested. Dynamic objects (*N*) are embedded in a home page, and requested after parsing the homepage. We set the static object request rate as $\lambda_1$ and the dynamic object request rate as $\lambda_2$ respectively. Figure 2 visualizes this case.

Figure 2 represents the Hyper-exponential distribution [11], which chooses the $i^{th}$ negative exponential distribution with the rate $\lambda_i$ and mean $1/\lambda_i$. That is, the density function is given by

$$f(S) = \sum_{i=1}^{2} p_i \lambda_i e^{-\lambda_i S} \qquad S \geq 0 \tag{7}$$

The $j^{th}$ moment is given by

$$E(S^j) = j! \sum_{i=1}^{2} \frac{p_i}{\lambda_i^j} \tag{8}$$

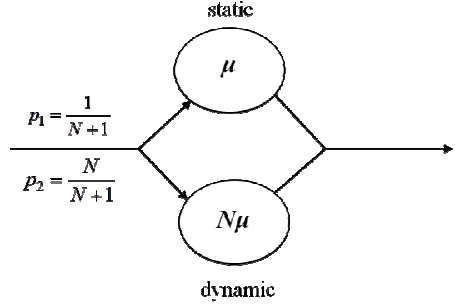

Figure 2. Graphical representation of web object requests

$E(S)$ and $E(S^2)$ are the first and the second moment of the web object service time, respectively. They are obtained by (9).

$$E(S) = \frac{2}{(N+1)\lambda} \qquad E(S^2) = \frac{2}{N\lambda^2} \tag{9}$$

By substituting $E(S)$ and $E(S^2)$ in (1), we obtain the mean queueuing delay for $M/H_2/1$ model in (10).

$$W_H = \frac{N+1}{\lambda(N-1)N} \tag{10}$$

## 3. MEAN WAITING TIME FOR MULTIPLE USERS

We first find out the number of simultaneous users satisfying that $M/H_2/1$ queueing delay($W_H$) is equal to the queueing delay in TDM($W_{TDM}$). From (6) and (10),

$$W_H = W_{TDM} \rightarrow m = \frac{2(1-\lambda)(N+1)}{\lambda(N-1)N} \tag{11}$$

Now, we consider the mean waiting time when $m$ concurrent users require access to a web object on a web server.

Web object is divided into multiple packets with a maximum segment size (MSS) by TCP in a transport layer. Let the web object size to be $\theta$, and MSS to be $mss$, the relation between the number of packets ($n$) and the web object size is given by (12).

$$n = \left\lceil \frac{\theta}{mss} \right\rceil \tag{12}$$

When multiple clients ($m$) request the same object, each user's expected service time ($E(S)$) is the same. However, exact finish time can vary due to the queueing delay; Clients must wait for the completion of the service.

We assume the asynchronous time division multiplexing based on packets for web service. When a client requests an object from the server, *n* packets are included in the object. E(*S*) means total response time that each client expects. Now, we assume that a packet based round robin scheduling policy is used in the multiple web access. This situation can be depicted in Figure 3.

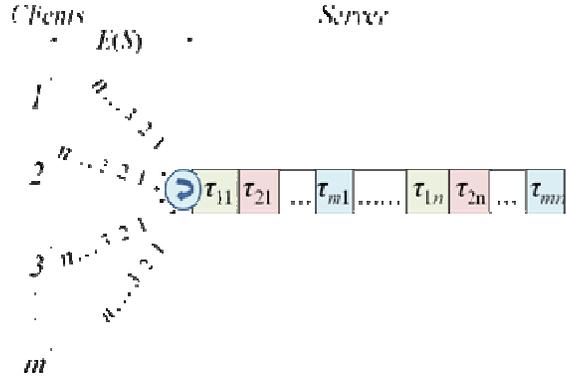

Figure 3. Scheduling for multiple web service

In Figure 3, $\tau_{ij}$ represents $j^{th}$ packet service time of the $i^{th}$ user. In order to simplify the modeling, we let $\tau_{ij} = \tau(\forall\ i, j)$, and then we can derive the mean waiting time as (13) [12].

$$W_R = \frac{1}{m}\sum_{i=1}^{m}\left[(m-i)\tau + m(m-1)(n-1)\tau\right]$$
$$= \frac{(m-1)(2n-1)\tau}{2} \quad (13)$$

Now, by assuming that the service time for single packet ($\tau$) is equal to one time slot in the steady state of the system, we can infer that Figure 1 and Figure 3 will be approximated.

Therefore, we can obtain the number of packets (*n*) satisfying that mean waiting time by (13) is equal to the mean queueing delay for M/D/1 with vacations model in the TDM by (6) as (14). Here, *m* represents the number of users (*m*) obtained by (11).

$$W_R = W_{TDM} \rightarrow \frac{(\underline{m}-1)(2n-1)}{2} = \frac{\underline{m}}{2(1-\lambda)}$$
$$\rightarrow n = \frac{(1-\lambda)(\underline{m}-1)+\underline{m}}{2(1-\lambda)(\underline{m}-1)} \quad (14)$$

In M/H$_2$/1 model, $\tau$= E(*S*)/*n*= 2/[*n*(*N*+1)$\lambda$] when every $\tau$ is same. Thus, the number of packets (*n*) satisfying that mean waiting time by (13) is equal to the mean queueing delay for M/H$_2$/1 by (10) is given by (15).

$$W_R = W_H \rightarrow \frac{(\underline{m}-1)(2n-1)}{2}\times\frac{2}{n\lambda(N+1)}$$
$$= \frac{\underline{m}}{2(1-\lambda)} \rightarrow n = \frac{(\underline{m}-1)(N-1)N}{2(\underline{m}-1)(N-1)N-(N+1)^2} \quad (15)$$

From (12), (14) and (15), we can obtain the web object size (*θ*) satisfying that mean waiting time for multiple web access environment is equal to the mean queueing delay for M/D/1 with vacations model and M/H$_2$/1 model, respectively.

$$\theta = \begin{cases} \dfrac{(1-\lambda)(\underline{m}-1)+\underline{m}}{2(1-\lambda)(\underline{m}-1)} \times mss & \text{for } W_R = W_{TDM} \\ \dfrac{(\underline{m}-1)(N-1)N}{2(\underline{m}-1)(N-1)N-(N+1)^2} \times mss & \text{for } W_R = W_H \end{cases} \quad (16)$$

In (16), for $W_R = W_{TDM}$, $\underline{m} \geq 2$ and $\lambda \leq 1$. For $W_R = W_H$, $\underline{m} > 1+(N+1)^2/2N(N-1)$.

## 4. PERFORMANCE EVALUATION

We first compute the number of users ($\underline{m}$) varying $N$ by using (11). Figure 4 represents the number of users when $mss = 1460$ B for various $\rho$ and $N$.

In Figure 4, when $\rho$ and $N$ are both small, the number of users is very large, but as $N$ increases, converges to 1. That is, the imbedded number of objects included in a web page can affect the simultaneous access number of users. The reason is why the number of users and the number of embedded objects in a page should be balanced in order to satisfy the given mean queueing delay. Although it is not presented in the figure, as $\rho$ increases, the number of simultaneous access users decreases to 1, so we cannot find out the web object size because a denominator has infinite value in (16).

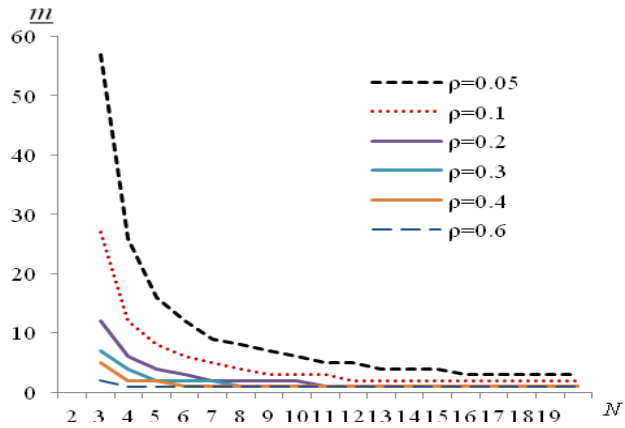

Figure 4. The number of users ($\underline{m}$) when $mss=1460$ varying $N$

Now, we compute the mean web object size ($\theta$) satisfying $W_R=W_{TDM}$ and $W_R=W_H$ respectively for varying $N$ given $\rho$ and $mss$. Table 1 shows the mean object size when $\rho=0.01$. Given $mss=1460$ B, mean object size is 1482 B for $W_R=W_{TDM}$ and 742 B for $W_R=W_H$. When $mss=536$ B, mean object size is 544 B for $W_R=W_{TDM}$ and 273 B for $W_R=W_H$. Therefore, the mean object size when $W_R=W_H$ is smaller than when $W_R=W_{TDM}$ regardless of $mss$.

Table 2 and Table 3 show the mean object size when $\rho=0.05$, $\rho=0.1$ and $\rho=0.2$, respectively. In Table 2 and Table 3, we find also that the mean object size when $W_R=W_H$ is smaller than when $W_R=W_{TDM}$ regardless of $mss$. That is, when the web service time has Hyper-exponential distribution, smaller object web size is required.

Table 1. Mean object size ($\theta$) satisfying $W_R=W_{TDM}$ and $W_R=W_H$ for varying $N$ when $\rho=0.01$

| N | $W_R=W_{TDM}$ | $W_R=W_H$ | $W_R=W_{TDM}$ | $W_R=W_H$ |
|---|---|---|---|---|
|  | $mss=1460$ | | $mss=536$ | |
| 2 | 1470 | 736 | 540 | 270 |
| 3 | 1473 | 738 | 541 | 271 |
| 4 | 1476 | 739 | 542 | 271 |
| 5 | 1480 | 741 | 543 | 272 |
| 6 | 1483 | 743 | 545 | 273 |
| 7 | 1487 | 745 | 546 | 274 |
| 8 | 1491 | 747 | 547 | 274 |
| 9 | 1495 | 749 | 549 | 275 |
| mean | 1482 | 742 | 544 | 273 |

We define the ratio of the mean object size satisfying $W_R=W_{TDM}$ over the mean object size satisfying $W_R=W_H$ as (17).

$$r = \frac{\theta_{W_R=W_{TDM}}}{\theta_{W_R=W_H}} \qquad (17)$$

Figure 5 depicts the ratio ($r$) by (17) for varying $\rho$. The size of $mss$ can not affect the ratio, but $\rho$ can affect it. The ratio is about 2 when $\rho=0.01$, however is decreased into 1 when $\rho=0.2$.

Table 2. Mean object size ($\theta$) satisfying $W_R=W_{TDM}$ and $W_R=W_H$ for varying $N$ when $\rho=0.05$

| N | $W_R=W_{TDM}$ | $W_R=W_H$ | $W_R=W_{TDM}$ | $W_R=W_H$ |
|---|---|---|---|---|
|  | $mss=1460$ | | $mss=536$ | |
| 2 | 1512 | 761 | 555 | 279 |
| 3 | 1529 | 771 | 561 | 283 |
| 4 | 1550 | 784 | 569 | 288 |
| 5 | 1568 | 795 | 576 | 292 |
| 6 | 1594 | 813 | 585 | 298 |
| 7 | 1608 | 819 | 590 | 301 |
| 8 | 1626 | 830 | 597 | 305 |
| 9 | 1652 | 848 | 607 | 311 |
| mean | 1580 | 803 | 580 | 295 |

Table 3. Mean object size ($\theta$) satisfying $W_R=W_{TDM}$ and $W_R=W_H$ for varying $N$ when $\rho=0.1$

| N | $W_R=W_{TDM}$ | $W_R=W_H$ | $W_R=W_{TDM}$ | $W_R=W_H$ |
|---|---|---|---|---|
|  | $mss=1460$ | | $mss=536$ | |
| 2 | 1572 | 799 | 577 | 293 |
| 3 | 1615 | 831 | 593 | 305 |
| 4 | 1657 | 858 | 608 | 315 |
| 5 | 1703 | 890 | 625 | 327 |
| 6 | 1744 | 917 | 640 | 337 |
| 7 | 1811 | 979 | 665 | 359 |
| 8 | 1947 | 1143 | 715 | 420 |
| 9 | 1947 | 1118 | 715 | 411 |
| mean | 1750 | 942 | 642 | 346 |

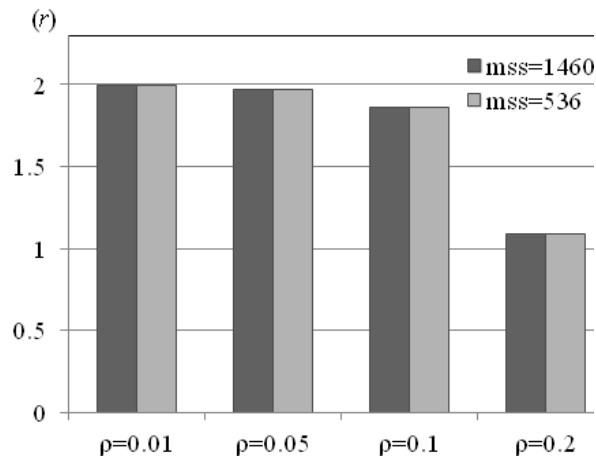

Figure 5. The ratio of mean object size satisfying $W_R=W_{TDM}$ over the mean object size satisfying $W_R=W_H$ for varying $\rho$

## 5. CONCLUSIONS

Mean object size in the multiple access environments is one of essential parameters to design and maintain the web service. To control the web object size is easy and very cheap maintenance method in order to satisfy the delay requirement of end-users. In this paper, we present an analytical model to estimate the web object size satisfying that the mean waiting time for multiple web service is equal to the mean queueing delay for the M/D/1 with vacations model in TDM system and the mean queueing delay for M/$H_2$/1 model, respectively. We first find out the number of users access web server simultaneously, and then derive the web object size models. Performance evaluations show that mean object size satisfying the M/D/1 with vacations model in TDM system is larger than that satisfying M/$H_2$/1 model, however the mean object size becomes the nearly same the utilization factor increases. Future works include more exact model applicable to the wire and wireless integrated network.

## REFERENCES

[1]     S. Ross, *Introduction to probability model*, Academic press, NewYork, 2010, p. 538, USA.
[2]     W. Shi, E. Collins, and V. Karamcheti, "Modeling Object Characteristics of Dynamic Web Content," *Journal of Parallel and Distributed Computing*, Elsevier Science, pp. 963-980, 1998.
[3]     R. Khayari, R. Sadre and B. R. Haverkort, "Fitting world-wide web request traces with the EM-algorithm, *Performance Evaluation*," Vol. 52, pp. 175-191, 2003.
[4]     A. Riska, V. Diev and E. Smirni, "Efficient fitting of long-tailed data sets into hyper-exponential distributions," *Proc. of IEEE Global Telecommunications Conference* (GLOBECOM 2002), Vol. 3, pp. 2513-2517, 2002.
[5]     Y. Lee, "Mean waiting delay for web service perceived by end-user in multiple access environment," *Natural Science* , vol. 2, Natural Science Institute of KNUE, pp. 55-58, 2012.
[6]     S. K. Bose, "M/G/1 with vacations," http:// www.iitg.ernet.in/skbose/qbook/Slide_Set_7.PDF, pp. 1-7, 2002.
[7]     N. Tian and Z. G. Zhang, *Vacation Queueing Model*, Springer Science and Business Media, pp. 10-11, 2006.
[8]     E. Modiano, "Communication systems engineering," MIT OpenCourseWare, http://ocw.mit.edu, pp. 1-19, 2009.
[9]     S. W. Fuhrmann, "Technical Note—A Note on the M/G/1 Queue with Server Vacations," *Operations Research*, Vol. 32, No. 6, pp. 1368-1373, 1984.
[10]    D. Bertsekas and R. Gallager, *Data Networks*, Prentice Hall, New Jersey, pp. 186-195, 2007.

[11]     M. S. Obaidat and N. A. Boudriga, *Fundamentals of Performance Evaluation of Computer and Tele-communication Systems*, Wiely, pp. 156-157, 2010.
[12]     Y. Lee, "Mean waiting time of an end-user in the multiple web access environment," *Proc. of the Sixth International Conference on Communication Theory, Reliability, and Quality of Service (CTRQ-2013)*, pp. 1-4, 2013.